\documentclass[12pt,a4paper]{article}

\begin{document}
\begin{flushright}
{\Large LAPTh--234/14}
\end{flushright}
\vskip 1cm
\begin{center}

{\Large\bf LanHEP - a package for automatic generation of Feynman rules from
 the Lagrangian. Updated version 3.2} \\[8mm]

{\large  A.~Semenov. }\\[4mm]

{\it  Joint Institute of Nuclear research, JINR, 141980 Dubna, Russia  }\\[4mm]
and \\
Laboratoire de Physique Th\'eorique LAPTh, Universit\'e de Savoie,\\
 Chemin de Bellevue, B.P. 110, F-74941 Annecy-le-Vieux, Cedex, France.\\[4mm]

\end{center}

 \begin{abstract}
 We present a new version 3.2 of the LanHEP 
 software package. New features include UFO output, color sextet particles
and new substutution techniques which allow to define new routines.
 \end{abstract}

\section*{Introduction}
The LanHEP program \cite{lanhep30} is developed for Feynman
rules generation from the Lagrangian. It reads
the Lagrangian written in a compact form, close to the one used in 
publications. It means that Lagrangian terms can be written 
with summation over indices of broken symmetries and using special symbols
for complicated expressions, such as covariant derivative and  
strength tensor for gauge fields. Supersymmetric theories can be described 
using the superpotential
    formalism and the 2-component fermion notation. The output
is Feynman rules in terms of physical fields and independent parameters
 in the form of CompHEP \cite{comphep} or CalcHEP \cite{calchep}
model files, which allows one to start calculations of processes in
the new physical model. Alternatively, Feynman rules can be generated
in FeynArts \cite{FeynArts} format or as LaTeX table. The program can also generate 
one-loop counterterms in the FeynArts format.

New version of the package can also generate Feynman rules in UFO \cite{UFO} format.
Use command line option {\tt -ufo} to select this format.

 The package can be downloaded from \\
 { \tt http://theory.sinp.msu.ru/\~{}semenov/lanhep.html}

\section{Color sextets}
 
One can use now particles belonging to 6-dimensional representation of the
color $SU(3)$ group, color sextets. LanHEP name for this representation  
is {\tt color c6} (and {\tt color c6b} for anti-sextets), it can be used
in the particle definition like:
\begin{quote} {\tt scalar s6/S6:('some sextet', mass M6=100, color c6).}
\end{quote}.

There are matrices for interaction of color sextets with triplets:
{\tt k\_c6} with two antitriplet indices and one with sextet index,
conjugated matrix {\tt k\_c6b} with two triplet and one antisextet indices,
and the matrix for interaction of sextets with gluon {\tt l\_c6} with
sextet, antisextet, and octet indices. So one can write the covariant 
derivative $\partial^\mu \delta_{ab} + ig\lambda^c_{6ab} G^\mu_c$
as {\tt deriv\^{}mu*delta(color c6) + i*g*l\_c6*G\^{}mu}.

Now color sextets are supported only for CalcHEP output. 

\section{Coefficients in the vertices}

A new feature in LanHEP allows to extract the expression from any generated vertex.
This can be used to implement the different contributions to the 
loop-induced $H\gamma\gamma$ and $Hgg$ vertices. 
The function which allows to extract a specific coupling is\\
 {\tt CoefVrt([ \sl particles \tt  ],[ \sl 
options \tt ])},\\
 where  {\sl particles} is the comma-separated list
of particles in the vertex, and {\sl options} specifies the tensor 
structure of the required coefficient as well as other options.  The only tensor structure that can be extracted are $1,\gamma^5,\gamma^\mu,\gamma^\mu\gamma^5, p^\mu $  and the function {\tt CoefVrt} must be called separately for each tensor structure. The second list can contain
the following symbols in any order:
\begin{itemize}
\item {\tt gamma} corresponds to only $\gamma^\mu$ matrices in the vertex, {\tt gamma5} to only  $\gamma^5$ matrices and 
{\tt gamma}, {\tt gamma5} to product of  $\gamma^\mu \gamma^5$ matrices. If none of these symbols are listed, then only 
terms which do not include $\gamma$-matrices will be extracted;

\item  {\tt moment( \sl P \tt )} where {\sl P} is a particle name selects 
the terms depending on this particle momentum, note that vertices corresponding to higher order operators and containing  the product of several momenta cannot be handled;
\item {\tt re, im} - take real or imaginary part of the expression. 
It can be combined with options above.
\item {\tt abbr} - use abbreviations in the returned expressions, this option is useful for long expressions.
\end{itemize}

For example, we can obtain expressions from the vertices 
of the Higgs boson interaction with sleptons (divided by the square of the slepton mass),
which contribute into $H\gamma\gamma$ vertex:
\begin{quote} \tt \_p=[eL,eR,mL,mR,l1,l2] in
    parameter AhS\_\_p=CoefVrt([anti(\_p),\_p,h]) /(mass \_p)**2/2.
\end{quote}

\section{Aliases}

Aliases allow to define substitution rules which are applied to all subsequent LanHEP
statements. For example, the statement {\tt alias mu=mue} allows to rename the 
parameter $\mu$  in the entire model. This statement must be placed before the declaration of
this parameter.

Aliases can have arguments, for example one can define electric charge for 
quarks:
\begin{quote}
 {\tt alias Q(u)=2/3, Q(c)=2/3, Q(t)=2/3, Q(d)=-1/3, Q(s)=-1/3, Q(b)=-1/3.}
\end{quote}

If a symbol appears on both the left-hand and the right-hand side of '=' sign
in the definition of an alias, then it denotes any term in the actual 
expression. For example,
\begin{quote} 
{\tt alias declScalar(name/antiname, massValue) = \\
\phantom{xxxxxxxxx} (scalar name/antiname:('scalar '\#name, mass M\#name=massValue)).}
\end{quote}
will declare a scalar particle named {\sl name} with mass $M_{name}$ equal to
{\sl massValue}. Here  \# sign means concatenation of the symbols to the left-hand and right-hand side.
If the symbol on the left-hand side of '=' sign begins with '\_' character, it denotes any
term. For example, {\tt alias Q(\_x)=0.} will define charge equal to zero for all
particles. Since aliases are applied to the expressions in the same order as they
were defined, more general definitions must follow more specific ones. On the other
hand, if a symbol which appears on both the left-hand and the right-hand side of '=' sign,
must denote only itself, it has to come on the left-hand side with unary plus operator:
{\tt +symbol}. Also unary plus in the rest of the model prevents its argument
from being processed by aliases.

As it was shown, aliases can define new statements to be used in the model. An alias
can be defined for a list of statements:
\begin{quote} {\tt
alias \sl newStat\tt = [(\sl Stat1\tt), (\sl Stat2\tt) ...].}
\end{quote}
In these statements new symbols (names for particles, parameters) can 
be generated by means of {\tt in} operator or by \# sign from the arguments
of {\sl newStat}. New symbols can also be defined in the separate {\tt local} statement
which creates temporary aliases used in the block of statements:
\begin{quote} {\tt alias declScalar(name/aname, massValue) = [\\
\phantom{xxxxxxxxx} (local sMass=M\#name), \\
\phantom{xxxxxxxxx} (scalar name/aname:('scalar '\#name, mass sMass=massValue)) ].}
\end{quote}

 \section*{Acknowledgments}
  This work was supported in part by  LIA-TCAP of CNRS and by the {\it
Investissements d'avenir}, Labex ENIGMASS.
This work  was  also supported by the Russian foundation for Basic Research, 
grant RFBR-12-02-93108-CNRSL-a.

 \end{document}